\documentclass[11pt]{article}

\usepackage[a4paper,margin=1in]{geometry}
\usepackage{setspace}
\usepackage{graphicx}
\usepackage{amsmath,amssymb}
\usepackage{hyperref}
\usepackage{titlesec}
\usepackage{enumitem}
\usepackage{tcolorbox}
\usepackage{caption}
\usepackage{cite}

\newtcolorbox{principlebox}[1]{
  colback=gray!5!white,
  colframe=black!75!black,
  fonttitle=\bfseries,
  title=#1
}

\title{\textbf{LERA: Reinstating Judgment as a Structural Precondition for Execution in Automated Systems}}

\author{
Jing (Linda) Liu \\
ORCID:0009-0002-1681-8563\\
Independent Researcher\\
\texttt{[linda@winston-battery.com]}
}

\date{January 11, 2026}

\usepackage[T1]{fontenc}
\usepackage{microtype}
\begin{document}

\maketitle

% -------------------------
% Abstract
% -------------------------
\begin{abstract}
As automated systems increasingly transition from decision support to direct execution, the problem of accountability shifts from decision quality to execution legitimacy. While optimization, execution, and feedback mechanisms are extensively modeled in contemporary AI and control architectures, the structural role of judgment remains undefined. Judgment is typically introduced as an external intervention rather than a native precondition to execution.

This work does not propose a new decision-making algorithm or safety heuristic, but identifies a missing structural role in contemporary AI and control architectures. This paper identifies this absence as a missing \emph{Judgment Root Node} and proposes \textbf{LERA (Judgment--Governance Architecture)}, a structural framework that enforces judgment as a mandatory, non-bypassable prerequisite for execution.

LERA is founded on two axioms: (1) execution is not a matter of system capability, but of structural permission, and (2) execution is not the chronological successor of judgment, but its structural consequence. Together, these axioms decouple execution legitimacy from computational capacity and bind it to judgment completion through a governance gate.

LERA does not aim to optimize decisions or automate judgment. Instead, it institutionalizes judgment as a first-class architectural component, ensuring that execution authority remains accountable. By reinstating judgment at the execution boundary, LERA establishes a foundational architecture for judgment-governed automation.
\end{abstract}


% =====================================================
\section{Introduction}
% =====================================================

As artificial intelligence and automated systems transition from decision support to direct execution, human society is entering a new technological phase. Systems are no longer confined to generating recommendations; they increasingly possess the capacity to trigger irreversible actions in the physical world, social systems, and critical infrastructures\cite{bostrom2014superintelligence}\cite{amodei2016concrete}. The widespread deployment of autonomous driving, intelligent power grids, financial automation, industrial control, and military systems has made execution itself a central locus of risk, responsibility, and governance.

Most contemporary discussions on AI safety and governance focus on model capability, alignment techniques, risk assessment, or ethical principles\cite{russell2019human}. Yet at the level of system architecture, a more fundamental and consequential question has remained largely unexamined: \textbf{what structural position does human judgment occupy in highly automated systems?}

In prevailing system designs, human judgment is typically introduced as an external intervention—such as approval workflows, manual review, exception handling, or emergency overrides. While these mechanisms appear to incorporate “human involvement” at an operational level, architecturally judgment remains outside the system core, functioning as an auxiliary input that can be delayed, bypassed, or formally substituted.

From an architectural perspective, this reveals a foundational structural vacancy. Optimization, execution, and feedback loops are explicitly modeled and continuously reinforced, \textbf{while judgment—as the source of execution authority—has been systematically excluded from the architectural core under prolonged efficiency-driven system evolution, resulting in a condition that can be described as architectural displacement.}

Judgment is not the generation of execution strategies, nor the comparison of alternatives for efficiency. It is a qualitative determination of whether execution should occur at all. When judgment is not institutionally embedded prior to execution, no amount of human intervention can prevent responsibility from being absorbed by system complexity as automation scales.

This paper identifies this absence as the missing \textbf{Judgment Root Node}. A Judgment Root Node denotes the native structural position that judgment must occupy within a system architecture: a position that must be explicitly satisfied before execution becomes possible, and that cannot be fulfilled by the system itself.

To address this structural vacancy, this paper introduces \textbf{LERA (Judgment–Governance Architecture)}. LERA is not designed to improve decision quality or expand system autonomy. Its purpose is to structurally enforce judgment as a prerequisite for execution and to bind execution legitimacy to the completion of judgment through a non-bypassable governance gate.

The contribution of this work lies not in proposing a new algorithm or management process, but in defining a structural role that has not been formally articulated in existing system architectures:
\textbf{judgment must occupy a non-bypassable position within the system, and execution legitimacy must originate from it.}

By formalizing the Judgment Root Node and establishing an irreducible structural relationship between judgment and execution, this paper seeks to provide a more foundational and enduring starting point for AI governance and automated system design than existing “human-in-the-loop” paradigms.

% =====================================================
\section{Problem Statement: The Missing Structural Role of Judgment}
% =====================================================

\subsection{Judgment Is Not Optimization and Not Execution}

In contemporary AI and automated system architectures, the concept of judgment is often implicitly conflated with either \textbf{optimization} or \textbf{decision execution}. Such equivalence is logically incorrect.

Optimization addresses the problem of computing improved outcomes under predefined objectives and constraints. Execution concerns the translation of a given decision into concrete, operational actions.

Judgment, however, does not operate within either of these domains.
Judgment does not answer the question of how to do something better, nor the question of how to execute.
Instead, it addresses a logically prior question:

\textbf{\textit{Whether the system should enter an executable state at all under the given circumstances.}}

This question neither involves objective function refinement nor action sequence generation.\\It concerns the legitimacy and permissibility of execution itself.

When judgment is misinterpreted as part of optimization, it is reduced to parameter tuning or objective reweighting.
When judgment is treated as part of execution, it degenerates into action selection or policy branching.

Both misinterpretations lead to the same outcome:
judgment is dissolved within system processes rather than established as a distinct structural role.This distinction allows us to examine judgment not functionally, but architecturally.

\subsection{The Structural Vacancy of Judgment in Contemporary System Architectures}

Once judgment is distinguished from optimization and execution, a more fundamental question emerges:

\textbf{\textit{Where does judgment reside within the system architecture?}}

From an architectural perspective, contemporary AI and automated systems explicitly model the following elements:
\begin{itemize}[noitemsep,topsep=2pt]
  \item objectives and constraints
  \item optimization processes
  \item decision generation
  \item execution pathways
  \item feedback and correction loops
\end{itemize}

Yet among these components, no dedicated structural position exists for judgment.

Judgment is neither modeled as a precondition to execution nor granted the authority to determine whether execution may proceed. Instead, it is typically placed at the system periphery, appearing as an external intervention or auxiliary mechanism.

This absence is not an implementation oversight but a \textbf{structural vacancy}.

While systems formalize how to execute and how to optimize, they fail to provide a native structural host for the question of whether execution should occur.
Execution remains structurally feasible by default, whereas judgment is not defined as a necessary prerequisite.

Thus, judgment is not absent in practice; it is absent in structure.
It never becomes part of the execution logic itself.

We refer to this phenomenon as the \textbf{structural vacancy of judgment}—the lack of a native architectural slot capable of hosting judgment as an execution precondition.

\subsection{Judgment Is Not Participation: Structural Priority and Non - Bypassability}

Human judgment is not absent from existing systems. On the contrary, many architectures incorporate human participation in various forms.

However, the effectiveness of judgment does not depend on whether humans participate, but on where judgment is positioned structurally.

Judgment constitutes a genuine governing function only if two conditions are satisfied:

\begin{enumerate}[label=\textbf{\arabic*.}, noitemsep]
  \item Judgment occurs \textbf{prior to execution}.
  \item Judgment is \textbf{structurally non-bypassable}.
\end{enumerate}

If judgment intervenes during execution or corrects outcomes afterward, it does not determine whether execution may occur; it merely influences execution behavior.

Likewise, if judgment is optional or skippable—even when labeled as “approval” or “review”—the system remains capable of progressing without it.

In such designs, execution is the default state, and judgment exists as an auxiliary step outside the execution path.

Therefore, the core issue is not participation but precedence:

\textit{Whether judgment is structurally defined as a mandatory, non-bypassable precondition for execution.}

Only under these conditions does judgment cease to be advisory input and become a governing structural element.

\subsection{Why Existing Paradigms Fail to Address the Structural Role of Judgment}

Multiple paradigms have been proposed to integrate judgment into system operation, most notably:

\begin{itemize}[noitemsep,topsep=2pt]
  \item Human-in-the-Loop
  \item Approval Workflows
  \item Safety Overrides / Emergency Stops
\end{itemize}

Although these paradigms introduce human factors, they share a common limitation:

\textit{None alters the default executability of the system.}

\textbf{Human-in-the-Loop (HITL)} incorporates human input into decision or execution processes, yet in most implementations, execution remains feasible even in the absence of human judgment\cite{sheridan1992telerobotics}. Judgment is informative rather than determinative.

\textbf{Approval workflows} rely on procedural compliance rather than structural enforcement\cite{parasuraman2000model}. Under automated, urgent, or large-scale conditions, approvals may be bypassed, downgraded, or retroactively recorded.

\textbf{Safety overrides} operate only after execution has begun and risk has emerged\cite{leveson2011engineering}. They are corrective, not preventive.

In all cases, judgment remains optional, procedural, or reactive.It is never established as a non-bypassable execution precondition.

Consequently, these paradigms fail to fill the structural vacancy of judgment.

\subsection{Structural Conclusion: Judgment Is Absent as an Execution Precondition}

The preceding analysis leads to an unavoidable conclusion:

\textit{In contemporary AI and automated system architectures, judgment is not established as a structural precondition for execution.}

Judgment appears in multiple forms—human input, approvals, interventions—but never as a governing structural constraint that must be satisfied before execution becomes possible.

Execution is typically treated as the system’s default state, while judgment functions as a supplement or correction\cite{astrom2010feedback}\cite{perrow1984normal}.

In other words, modern systems operate under a \textbf{Default-Execute} assumption: execution is structurally permitted unless explicitly interrupted.

LERA introduces a fundamental reversal of this assumption.
It shifts the system’s default state from \textbf{Default-Execute} to \textbf{Default-Block}, under which execution is structurally infeasible until judgment is satisfied.

\subsection{Defining the Structural Vacancy: The Absence of a Judgment Root Node}

This long-standing but previously unnamed deficiency can now be formally defined.

The structural vacancy consists of the absence of a \textbf{non-bypassable, execution-prior structural position capable of hosting judgment.}

We designate this missing position as the absence of a \textbf{Judgment Root Node}.

The Judgment Root Node does not denote a specific algorithm, interface, or workflow. It denotes a \textbf{structural role} that determines whether a system is eligible to enter an executable state.

From a physical systems perspective, this role is analogous to a \textbf{circuit breaker}.Unless the breaker is closed, current cannot flow; likewise, unless the Judgment Root Node is satisfied, execution is structurally impossible.

Judgment thus ceases to be advisory or corrective and becomes a physical precondition of execution.

When this structural node is absent, judgment—regardless of its form—remains external and non-binding.

The \textbf{LERA (Linda Energy Reliability Architecture)} is proposed as an architectural response to this structural vacancy. Its purpose is not to optimize decisions or enhance execution, but to provide a native, non-bypassable structural host for judgment.

Subsequent sections will formalize this architecture and demonstrate how the Judgment Root Node is established as a root-level governance condition for execution.

% =====================================================
\section{The LERA Architecture: Institutionalizing Judgment as a Structural Precondition}

\subsection{Design Objective: Making Judgment Structurally Effective}
Section 2 established that contemporary AI and automated systems do not lack judgment per se; rather, they lack \textbf{a structural position in which judgment becomes effective}.

The objective of \textbf{LERA (Linda Energy Reliability Architecture)} is therefore not to introduce additional sources of judgment, nor to improve decision quality or execution efficiency. Instead, LERA addresses a more fundamental architectural problem:

\textit{How can judgment be made a necessary structural precondition for execution?}

Accordingly, LERA should not be understood as an optimization framework or a risk assessment tool. It is a \textbf{governance architecture} whose central purpose is to transform judgment from an external intervention into an internal structural condition that determines executability.

\subsection{Architectural Principle: Structural Separation of Judgment and Execution}
The first foundational principle of LERA is the \textbf{structural separation of judgment and execution}.

In many existing systems, judgment, decision-making, and execution are compressed into a single logical pipeline\cite{wiener1948cybernetics}. Judgment is either algorithmically subsumed into optimization or procedurally diluted within execution workflows, thereby losing its authority over whether execution may occur.

LERA explicitly requires that:

\textit{Judgment formation and execution authorization be structurally separated.}

This separation is not merely temporal. It establishes an \textbf{asymmetric dependency}:
\begin{itemize}[noitemsep,topsep=2pt]
  \item Judgment may exist without execution.
  \item Execution, however, cannot exist without judgment.
\end{itemize}

By enforcing this asymmetry, LERA ensures that judgment is not consumed by execution logic and that its structural authority is preserved.
\subsection{LERA-J: The Judgment Formation Layer}
Within the LERA architecture, the formation of judgment is assigned to a dedicated structural layer termed \textbf{LERA-J (Judgment Layer)}.

The function of LERA-J is not to compute optimal actions or to generate executable commands. Instead, it performs a \textbf{qualitative judgment of the operational context}, producing a determination regarding whether execution may be permitted.

Key characteristics of LERA-J include:

\begin{itemize}[noitemsep,topsep=2pt]
  \item It may integrate heterogeneous inputs, including algorithmic analysis, rule-based reasoning, and human judgment.
  \item Its output is not an action or policy, but a \textbf{judgment signal} concerning executability.
  \item It possesses no execution capability and exercises no direct control over physical or logical actions.
\end{itemize}

This design ensures that judgment formation remains conceptually and structurally independent from execution pathways.

\subsection{LERA-G: The Governance Gate as a Non-Bypassable Structural Interlock}
For judgment to function as a true precondition to execution, it must not only be formed but also be \textbf{structurally enforced}.

LERA achieves this enforcement through \textbf{LERA-G (the Governance Gate)}.

Crucially, LERA-G is \textbf{not} a procedural approval step, a conditional branch, or a soft authorization mechanism.
It is neither an if–else construct nor a workflow abstraction that can be simulated or bypassed through alternative logic paths.

From an architectural standpoint, \textbf{LERA-G constitutes a primitive structural component.}
Its role is to operate as a \textbf{structural interlock} that maintains the system in a state of \textbf{execution-quiescence} until a valid governance condition is satisfied.

Within the LERA architecture:
\begin{itemize}[noitemsep,topsep=2pt]
  \item All execution pathways are structurally routed through LERA-G.
  \item Prior to LERA-G closure, execution channels remain inert.
  \item Any attempt to initiate execution while bypassing LERA-G is rendered structurally invalid.
\end{itemize}

This constraint is not procedural but hard-wired—a non-bypassable hard constraint embedded at the execution boundary.
\subsubsection{Asymmetry and Non-Substitutability}
To prevent LERA-G from being reduced to a generic conditional trigger, the architecture explicitly requires that \textbf{LERA-G closure depend on an asymmetric governance signal with verifiable structural properties}.

Specifically:
\begin{itemize}[noitemsep,topsep=2pt]
  \item Activation of the execution pathway cannot be triggered by arbitrary logical conditions.
  \item Execution authorization requires an \textbf{asymmetric signal coupling} generated exclusively by \textbf{LERA-J} and validated by LERA-G.
\end{itemize}

This design ensures that no functionally equivalent substitute—procedural, logical, or symbolic—can activate execution.

In structural terms, LERA establishes a strict \textbf{causal dependency} between judgment and execution:

\begin{itemize}[noitemsep,topsep=2pt]
  \item \textbf{LERA-J} performs an \textbf{ontological qualitative determination} of the operational context.
  \item \textbf{LERA-G} authorizes execution solely upon receipt of a \textbf{validated governance signal} originating from LERA-J.
  \end{itemize}

In the absence of such a signal, any downstream command—regardless of computational readiness or decision-generation intensity—is rendered into a state of \textbf{operational infeasibility}.

\subsubsection{Governance Function as a Physical / Logical Entity}
Functionally, LERA-G implements governance.

Structurally, however, it behaves more like an industrial safety interlock than an administrative approval mechanism.

By defining LERA-G as a \textbf{physically or logically decoupled interlock mechanism}, LERA ensures that:

\begin{itemize}[noitemsep,topsep=2pt]
  \item Judgment absence cannot be masked by increased model intelligence or optimization strength.
  \item Execution cannot be forced by escalation within the decision-generation layer.
  \end{itemize}
  
In this sense, LERA-G does not merely enforce governance rules;
it \textbf{defines the boundary of executability itself}.

Execution, under LERA, is no longer a default system capability.

It becomes a conditionally unlocked state, made possible only through the structural closure of LERA-G.
\subsection{The Judgment Root Node in LERA}
Through the coordinated operation of \textbf{LERA-J} and \textbf{LERA-G}, judgment acquires a definitive structural position within the LERA architecture.

This position is what Section 2 defined as the \textbf{Judgment Root Node}.

The Judgment Root Node is not identical to LERA-J, nor is it reducible to LERA-G. Rather, it is the \textbf{structural condition jointly constituted} by both:

\begin{itemize}[noitemsep,topsep=2pt]
  \item LERA-J forms judgment.
  \item LERA-G enforces its structural validity.
  \end{itemize}

Only when judgment is produced by LERA-J and accepted by LERA-G does the system become structurally eligible to enter an executable state.

In this sense, the Judgment Root Node functions as:

\textit{the sole legitimate structural interface between judgment and execution.}
\subsection{Structural Implication: Judgment as a Root-Level Governance Condition}
By fixing judgment at the level of a Judgment Root Node, LERA effects a fundamental architectural shift.

Judgment is no longer:
\begin{itemize}[noitemsep,topsep=2pt]
  \item advisory input,
  \item a procedural checkpoint, or
  \item a post-hoc corrective mechanism.
  \end{itemize}

Instead, it becomes a \textbf{root-level governance condition}.

Under LERA, execution is no longer the system’s default state. It is a \textbf{conditionally unlocked state}, attainable only through satisfaction of the Judgment Root Node.

This shift does not depend on increased model sophistication, tighter optimization, or more elaborate workflows. It arises solely from a redefinition of the system’s structural dependencies.

\subsection{Scope and Intent}
LERA does not attempt to specify the substantive content of judgment, nor does it seek to replace human decision-makers or domain experts.

Its concern is not \textbf{whether judgment is correct}, but whether judgment is \textbf{structurally effective}.

The quality of judgment is an optimization problem.

The authority of judgment is a governance problem.

LERA addresses the latter by providing a non-bypassable structural foundation through which judgment becomes operative.

% =====================================================
\section{Fundamental Axioms of LERA}
% =====================================================
This section presents the foundational axioms that constitute the core of the LERA (Judgment–Governance Architecture).
These axioms do not prescribe implementation details, optimization strategies, or the technical nature of judgment agents. Instead, they define the non-negotiable structural relationship between judgment and execution in any system capable of producing irreversible effects in the physical or social world.

\begin{principlebox}{Axiom 1: The Principle of Execution Infeasibility}
\textbf{Execution is not a matter of system capability, but a matter of structural permission.}

In the absence of verified governance authorization through LERA-G, execution is structurally infeasible, regardless of computational capacity, decision quality, or system intelligence. Even when a system possesses the technical means to act, the execution pathway remains architecturally inaccessible until the governance gate is explicitly closed.

This axiom establishes a fundamental constraint: execution is not an operational continuation of computation, but a structurally gated transition. LERA-G therefore does not function as a procedural checkpoint or a post-hoc approval mechanism. It is a non-bypassable structural interlock that maintains the system in a state of execution quiescence until the required governance conditions are satisfied.

By defining execution infeasibility at the architectural level, this axiom eliminates all forms of soft approval, deferred authorization, or logical circumvention. Execution cannot be “temporarily blocked” or “conditionally allowed”; it is either structurally permitted or structurally impossible.
(Formally, execution E is defined only if governance authorization G(J) has been satisfied; otherwise, E is structurally undefined.)
\end{principlebox}

\begin{principlebox}{Axiom 2: The Structural Precedence of Judgment}
\textbf{Execution is not the chronological successor of judgment, but its structural consequence.}

Judgment precedes execution not as a matter of process ordering, but as a condition of architectural legitimacy. Judgment is distinct from optimization, strategy selection, or decision generation. It is a qualitative determination of whether execution should occur at all.

Within the LERA architecture, execution does not merely follow judgment in time; it emerges from judgment as a causally dependent state transition. Without the completion of judgment, no legitimate execution pathway exists within the system architecture. Execution is therefore not delayed in the absence of judgment—it is undefined.

This axiom rejects all paradigms in which judgment is reduced to advisory input, auxiliary approval, or exceptional override. Instead, judgment is established as a first-class structural precondition for execution, without which execution cannot acquire legitimacy or architectural existence.

Together, these two axioms form the minimal and complete core of LERA:
\textbf{execution capability is decoupled from execution legitimacy, and legitimacy is structurally grounded in judgment.}
\end{principlebox}

% =====================================================
\section{Scope and Non-Goals}
% =====================================================

To prevent conceptual dilution and misinterpretation, this section clarifies the scope of LERA and explicitly states what the architecture does not attempt to solve.

\subsection{LERA Is Not an Optimization or Decision-Generation Framework}
LERA does not generate decisions, compare alternatives, or optimize outcomes. It does not evaluate efficiency, cost, risk, or performance metrics, nor does it participate in strategy selection or policy synthesis.

The purpose of LERA is not to improve decision quality, but to define the structural conditions under which execution becomes legitimate. It does not answer the question of what should be done better, but rather whether execution should be allowed to occur at all.
\subsection{LERA Does Not Define the Technical Nature of the Judgment Agent}
LERA does not prescribe who or what performs judgment, nor does it define the technical form, intelligence level, or cognitive properties of the judgment agent. It does not attempt to resolve philosophical or scientific debates regarding consciousness, personhood, or machine cognition.

Instead, LERA requires that any entity authorized to unlock execution must occupy an explicit, non-bypassable judgment position within the system architecture. The identity and form of such an entity are matters of governance choice rather than architectural necessity.
\subsection{Judgment Is Institutionalized, Not Automated}
LERA does not automate judgment, nor does it delegate judgment to the internal logic of the system itself. Its central function is to institutionalize judgment as a mandatory, explicit precondition for execution.

In the current socio-technical context, the capacity to bear responsibility for irreversible consequences is necessarily anchored in human agents or human-governed institutions. LERA is designed to ensure that this responsibility cannot be diluted, transferred, or implicitly absorbed by increasingly autonomous systems.

By structurally separating judgment from execution, LERA prevents responsibility from being dissolved into system complexity or obscured by automation layers.

LERA may introduce execution latency by requiring explicit judgment–governance closure prior to action. This work treats such latency not as a system inefficiency, but as a deliberate architectural trade-off: in high-risk domains involving irreversible consequences, structural safety and accountability take precedence over raw execution speed. Efficiency optimization remains a secondary concern once execution legitimacy is established.In this sense, LERA frames execution latency as a necessary cost of preserving human sovereignty at the boundary of irreversible action.
\subsection{Applicability Under Future Intelligence Paradigms}
LERA does not govern the evolution of intelligence or cognition. It governs the authorization boundary of execution.

Even under future scenarios involving advanced artificial intelligence or novel cognitive substrates, as long as execution entails irreversible effects, the requirement for a structurally explicit judgment precondition remains valid. Any system permitted to trigger such execution must be explicitly authorized at the governance level and bear the associated responsibility.
\subsection{Summary}
By clearly defining its scope and non-goals, LERA focuses on a single foundational problem:

\textbf{how to preserve accountable judgment at the boundary of execution in increasingly automated systems.}
LERA is not intended to increase system autonomy.
LERA serves as the final structural anchor for human sovereignty at the boundary of irreversible action.

% =====================================================
\section{Conclusion}
% =====================================================

To address this problem, this paper introduced LERA (Judgment–Governance Architecture) and articulated two foundational axioms that define an irreducible structural relationship between judgment and execution. LERA does not attempt to optimize outcomes or prescribe the technical nature of judgment agents. Instead, it establishes judgment as the structural source of execution legitimacy and enforces this relationship through a non-bypassable governance gate.

Within the LERA framework, execution is no longer a natural extension of system capability, but a structurally authorized transition. Judgment is no longer an advisory input, approval step, or exceptional override, but a mandatory institutional condition that must be satisfied before execution can occur.

In the current socio-technical context, the capacity to bear responsibility for irreversible consequences remains anchored in human agents and human-governed institutions. LERA is explicitly designed to prevent this responsibility from being diluted, transferred, or silently absorbed by increasingly autonomous systems.

\textbf{LERA serves as the final structural anchor for human sovereignty at the boundary of irreversible action.}

This position does not reject technological progress, but clearly constrains the boundary of execution. As long as systems retain the ability to produce irreversible effects, the structural requirement for judgment as a precondition to execution must persist. LERA does not offer a closed technical solution, but establishes a non-negotiable governance threshold that cannot be bypassed.

\bibliographystyle{plain}
\bibliography{references}

% =====================================================
\appendix
% =====================================================

\section{Appendix: Terminology and Conceptual Clarifications}
This appendix provides concise clarifications of key terms used throughout this paper. The purpose is not to expand the theory, but to prevent ambiguity, reinterpretation, or dilution of the structural concepts introduced.

\begin{description}[leftmargin=1.5cm]

\item[Judgment]
refers to a qualitative, context-sensitive determination regarding whether execution ought to occur. It is not reducible to optimization, prediction, or decision selection. Judgment concerns legitimacy and permissibility, not efficiency or utility.

\item[Decision]
Decision denotes the selection of an option among alternatives within a predefined solution space. Decisions may be optimized, automated, or statistically derived. Judgment, by contrast, operates prior to and outside this solution space.

\item[Execution]
is the act of committing an action into the physical or irreversible domain. Once execution occurs, its consequences may no longer be retractable or fully correctable.

\item[Judgment Root Node]
is a structural position within a system architecture where judgment must be completed before execution becomes possible. It is not a procedural step, advisory input, or post-hoc intervention, but a causal precondition to execution.

\item[Structural Precondition]
is a condition that must be satisfied at the architectural level for a system transition to occur. Unlike procedural checks, structural preconditions cannot be bypassed without altering the system’s architecture itself.

\item[LERA-J](Judgment Formation Layer)
denotes the layer responsible for forming judgment. It performs an ontological qualitative determination of the operational context and issues a root-level authorization signal. LERA-J does not execute actions.

\item[LERA-G](Governance Gate)
is a non-bypassable execution gate that enforces the structural dependency between judgment and execution. It functions as a structural interlock that maintains the system in an execution-inert state until a validated governance signal is received.

\item[Execution Infeasibility]
refers to a state in which execution is structurally impossible, regardless of system capability or intent. In LERA, execution infeasibility is the default condition in the absence of LERA-G closure.

\item[Structural Interlock]
is a physical or logical mechanism that prevents execution unless specific structural conditions are met. It differs from software-based conditional logic in that it cannot be bypassed through internal decision pathways.

\item[Default-Block]
Default-Execute describes architectures in which execution is assumed reachable unless explicitly stopped. 

Default-Block describes architectures in which execution is structurally blocked unless explicitly permitted through judgment completion.Default-Block: An architectural paradigm where execution is rendered structurally infeasible as the starting state.

\end{description}

\section{Appendix: Architectural Scope Illustration}
\textbf{Figure Caption}:The LERA architecture institutionalizes human responsibility by mandating the separation of judgment formation (J) and execution authorization (G).\textbf{Human judgment becomes effective only through LERA-G's enforcement of structural permission}.This ensures that any high-stakes operation on the physical world has an explicit, non-bypassable human anchor point for accountability.LERA-G appears both as a structural component of the architecture and as an operational gate instantiated during execution. These refer to the same governance function at different abstraction levels.
\begin{figure}[!ht]
    \centering
    \includegraphics[width=0.7\textwidth]{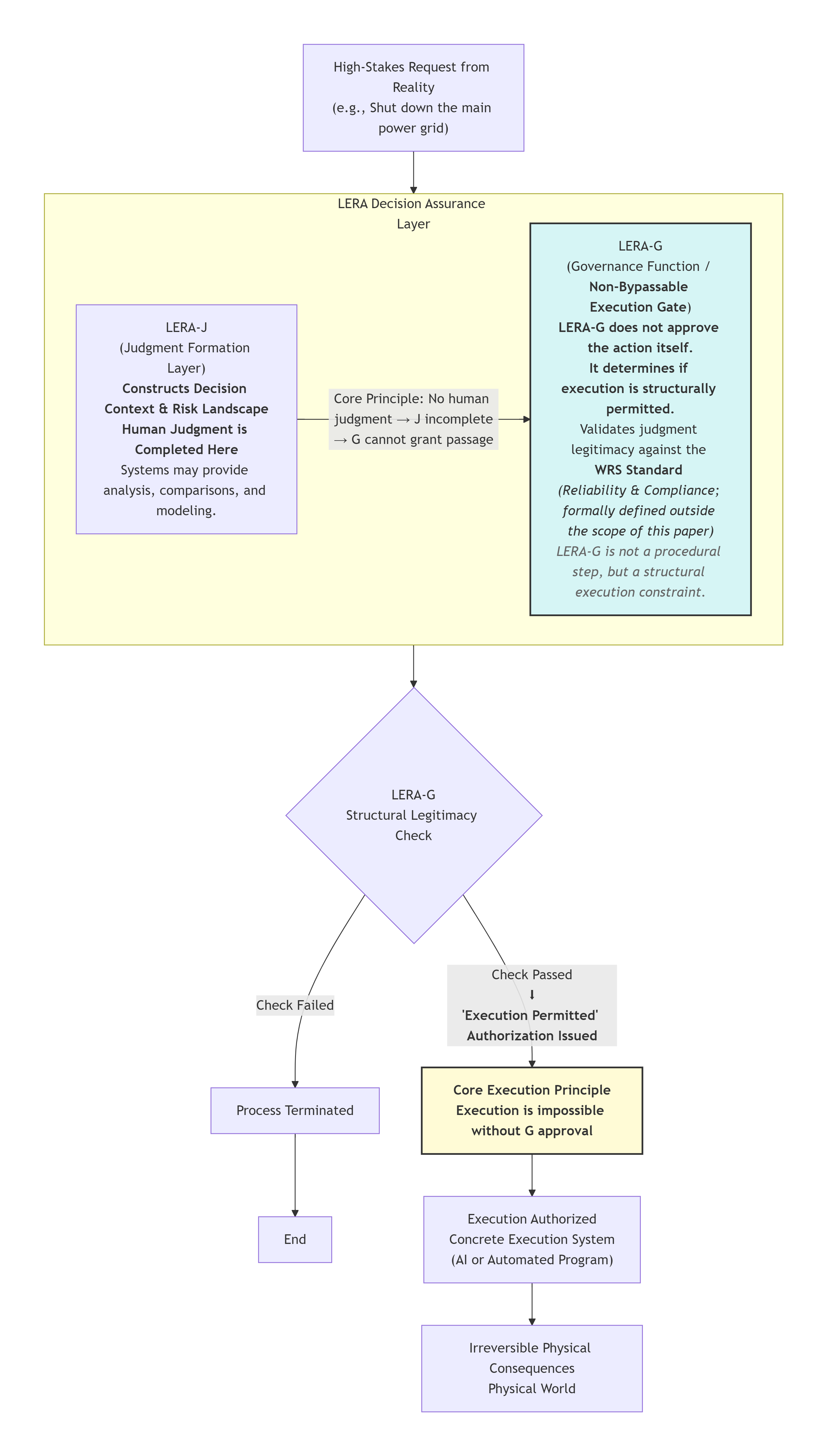}
    \caption{LERA Architecture Scope Illustration}
    \label{fig:lera_architecture} 
\end{figure}

\end{document}